\documentclass[useAMS,usenatbib]{mnras}
\usepackage{graphicx,psfig,cite,epsf}  
\usepackage{times}
\usepackage{subfigure}
\usepackage{textcomp}
\usepackage{multirow}
\usepackage{amsmath}	
\usepackage{amssymb}	
\usepackage{gensymb}
\usepackage{siunitx}

\def\apj{ApJ}                 
\def\apjl{ApJ}                
\def\apss{Ap\&SS}             
\def\aap{A\&A}                
\def\mnras{MNRAS}             
\def\pasp{PASP}               

\def\xmm{{\it XMM-Newton}}

\def\pdot {\dot P}
\def\nudot {\dot \nu}

\def\msun{{\rm M}_{\odot}}
\def\rsun{R_{\odot}}

\def\ltsima{$\; \buildrel < \over \sim \;$}
\def\lsim{\lower.5ex\hbox{\ltsima}}
\def\gtsima{$\; \buildrel > \over \sim \;$}
\def\gsim{\lower.5ex\hbox{\gtsima}}
\def\hd {HD\,49798}
\def\rx {RX J0648.0--4418}
\def\hr {HD\,49798/RX\,J0648.0--4418}

\title[HD 49798]{New X-ray observations of the hot subdwarf binary \hr\ }

\author[]{S. Mereghetti$^1$\thanks{E-mail: sandro.mereghetti@inaf.it}, F.~Pintore$^2$, T.~Rauch$^3$, N.~La Palombara$^1$, P.~Esposito$^4$, S.~Geier$^5$,  
\newauthor I.~Pelisoli$^{5,6}$, M.~Rigoselli$^1$, V.~Schaffenroth$^5$,  A.~Tiengo$^4$ \\
$^1$ INAF -- IASF Milano, Via E. Bassini 15, 20133 Milano, Italy \\
$^2$ INAF -- IASF Palermo, Via U. La Malfa 153, 90146 Palermo, Italy \\
$^3$  Institute for Astronomy and Astrophysics,  Kepler Center for Astro and Particle Physics,  Eberhard Karls University, Sand 1, 72076 T\"ubingen, Germany\\
$^4$ Scuola Universitaria Superiore IUSS Pavia, Piazza della Vittoria 15, 27100 Pavia, Italy\\
$^5$ Institut f\"{u}r Physik und Astronomie, Universit\"{a}t Potsdam, Haus 28, Karl-Liebknecht-Str. 24/25, D-14476 Potsdam-Golm, Germany\\
$^6$  Department of Physics, University of Warwick, Coventry, CV4 7AL, UK
}

\date{Accepted 2021 April 07. Received 2021 April 07; in original form 2021 February 23}

\pubyear{2021}

\begin{document}

\maketitle

\begin{abstract}

\hr\  is the only confirmed X-ray binary in which the mass donor is a hot subdwarf star of O spectral type and, most likely, it contains a  massive white dwarf  (1.28$\pm$0.05 $\msun$) with a very fast spin period of 13.2 s.
Here we report the results of new \xmm\ pointings of this peculiar  binary, carried out in 2018 and in 2020, together with a reanalysis of  all the previous observations.
The new data indicate that the compact object is still spinning-up  at a steady rate of $(-2.17\pm0.01)\times10^{-15}$ s s$^{-1}$, consistent with its interpretation in terms of a young contracting white dwarf.   Comparison of observations obtained at similar orbital phases, far from the ecplise,  shows  evidence for long term variability of the hard ($>$0.5 keV) spectral component at a level of $\sim$(70$\pm$20)\%, suggesting the presence of time-dependent inhomogeneities in the weak stellar wind of the \hd\ subdwarf.  
To   investigate better the soft spectral component that dominates the X-ray flux from this system, we
computed a theoretical model for the thermal emission expected from an atmosphere with element abundances and surface gravity  appropriate for this massive white dwarf. This model gives a best fit with effective temperature of T$_{\rm eff}$=2.25$\times$10$^5$ K and an emitting area with radius of $\sim$1600 km, larger than that found with blackbody fits. This model also predicts a contribution of the pulsed emission from the white dwarf in the optical band significantly larger than previously thought and possibly relevant for   optical variability studies of this system.

\end{abstract}
\begin{keywords}
X-rays: binaries -- white dwarfs -- subdwarfs.
\end{keywords}

\section{Introduction}

\hr\ is the only known accreting X-ray  binary in which the mass donor belongs to the class of hot subdwarf stars \citep{heb16}.   
This binary, likely  the outcome of a common-envelope   stage,  is relevant in the context of evolutionary models of intermediate mass stars, possibly leading to the formation of millisecond pulsars or type Ia supernovae \citep{wan10,bro17,wu19}. 
It is also interesting because, through the study of its X-ray emission,  it offers the possibility to obtain some information on the weak stellar wind of the hot subdwarf   \citep{mer16r,krt19}.
 
\hr\ is composed of  a compact object   spinning at P=13.2 s and a  subdwarf star of O  spectral type  \citep{kud78,isr97,mer11}. The orbital period, determined through optical spectroscopy since   early observations of \hd,  is   1.55 days \citep{tha70}. 
The masses of the two stars are well constrained by the measurement of the optical and X-ray mass functions,
the system inclination being derived from the duration of the X-ray eclipse: the compact object has a mass of 1.28$\pm$0.05 $\msun$ and  is most likely a white dwarf, while the mass of \hd\   is 1.50$\pm$0.05 $\msun$ \citep{mer09}.  
The most recent  parallax    obtained with Gaia EDR3   corresponds to a distance of  521$\pm$14 pc \citep{gaia20}.

Contrary to the majority of X-ray binaries, which are transient or highly variable, \rx\ is a  persistent X-ray source: a  luminosity of $\sim$10$^{32}$ erg s$^{-1}$ was seen in all observations, that  now  span almost   thirty years.
With a radius of  $\sim$1$\rsun$,  \hd\ underfills its Roche lobe,  but it is one of the few hot subdwarfs with strong evidence for a stellar wind \citep{ham81,ham10}. The most recent estimate, obtained through the modeling of high-resolution UV/optical spectra, yields a mass loss rate of 2.1$\times10^{-9}$ $\msun$ yr$^{-1}$  and a wind terminal velocity of 1570 km s$^{-1}$ \citep{krt19}.
With these stellar wind parameters, the observed X-ray luminosity is consistent with that expected from accretion onto a massive white dwarf \citep{mer09,mer11}.  

Besides the low X-ray luminosity, much smaller than that expected if the accreting object were a neutron star, there are other arguments that support the presence of a white dwarf in this system.  
Most of the X-ray flux is emitted in a very soft and strongly pulsed thermal component, well fitted by a blackbody with temperature of  kT$\sim30$ eV and an emitting radius of $\sim$30  km \citep{mer16}. 
These dimensions are consistent with those of a hot spot on the surface of a  white dwarf,  but are too large for a neutron star. This emitting radius could be reconciled with a neutron star only if the  X-ray emission comes from the whole (or a large part of the) star surface, but this is at variance with the very high pulsed fraction of the soft X-ray pulse profile ($\sim$65\%).   

The   steady long-term spin-up of \rx  , at a rate of   $\pdot = -2.15\times10^{-15}$ s s$^{-1}$, is difficult to explain by accretion torques \citep{mer16}. On the other hand, \citet{pop18} showed that it is   consistent with that caused by the radial contraction of a  young white dwarf  with age of  a few millions years. Such a small age is in agreement with the evolutionary models that account for the properties of this binary. Finally, we note that population synthesis simulations of hot subdwarf binaries predict that those hosting white dwarfs  greatly outnumber those with a neutron star  \citep{yun05,wu18}.

To continue our long term monitoring of this unique X-ray binary,  we observed it   with \xmm\ in  2018 and 2020.   Here we present the results of these new observations, complemented by a reanalysis of all the previous data, including the first spectral fits with a white dwarf atmosphere model applied to this system.

\section{Data analysis and results}

\xmm\ observed \rx\ for $\sim$34 ks on 2018 November 8 and for $\sim$37 ks on 2020 February 27.  We use the data obtained with the EPIC instrument, that consists of one pn and two MOS cameras \citep{str01,tur01}. 
As in all the previous observations, the pn and MOS cameras were operated in  Full-Frame mode, giving time resolutions of 73 ms and 2.6 s, respectively.

To  compare properly  the new observations with the previous ones, we processed all the available \xmm\ data using SAS v.16.1 and the most recent calibration files. We filtered out time intervals with high background and selected single- and double-pixels events for the pn (PATTERN$\leq$4), and single- and multiple-pixels events for the MOS (PATTERN$\leq$12). The resulting exposure times are given in Table~\ref{log}.  Source and background events were extracted from circular regions with radii of 30'' and 60'', respectively. 

For the timing analysis we used also the ROSAT data obtained on  1992 November 11.
We selected the counts obtained with the PSPC instrument in the energy range  0.1--0.5 keV,  from a source region with radius of 1 arcmin. 

For all data sets, the times of arrival of  the events were corrected to the solar system barycenter  using the source coordinates given in Table~\ref{tim_par}.

\begin{table}
\center
\caption{Log of the \xmm\ observations of HD 49798.
\label{log}}
\scalebox{0.8}{\begin{minipage}{24cm}
\begin{tabular}{llcccc}
\hline
Date & Obs. ID &   Start & Stop              & Exposure   & Orbital phase\\
       &              & (MJD)  & (MJD)            & pn/MOS (ks)  &          \\
\hline
2002 May 03  & 0112450301 &   52397.46 & 52397.55 & 4.5/7.2  & 0.45 -- 0.48    \\
2002 May 04 & 0112450401 &   52397.98 & 52398.06 & 1.4/5.4  & 0.78 -- 0.81\\
2002 May 04 & 0112450501 &   52398.56 & 52398.59 & 0.6/2.5 & 0.17 -- 0.16\\
2002 Sep 17 & 0112450601 &   52534.58 & 52534.72 & 6.9/11.9 & 0.06 -- 0.11\\
2008 May 10 & 0555460201 &   54596.90 & 54597.38 & 36.7/43.0 & 0.56 -- 0.87\\
2011 May 02 & 0671240901 &   55683.55 & 55683.76 & 17.0/18.3 & 0.70 -- 0.82\\
2011 Aug 18 & 0671241001 &   55791.88 & 55792.07 & 15.0/16.4 & 0.69 -- 0.80\\
2011 Aug 20 & 0671241101 &   55793.46 & 55793.62 & 11.8/14.1 & 0.72 -- 0.80\\
2011 Aug 25 & 0671241201 &   55798.04 & 55798.27 & 18.0/19.2 & 0.67 -- 0.81\\
2011 Sep 03 & 0671241301 &   55807.35 & 55807.54 & 15.0/16.4 & 0.69 -- 0.80\\
2011 Sep 08 & 0671241401 &   55811.10 & 55812.19 & 15.0/16.4 & 0.70 -- 0.80\\
2013 Nov 10 & 0721050101 &   56605.80 & 56606.25 & 37.9/39.1 & 0.60 -- 0.86\\
2014 Oct 18 & 0740280101 &   56948.37 & 56948.71 & 27.1/29.1 & 0.95 -- 0.15\\
2018 Nov 08 & 0820220101 &   58430.46 & 58430.92 & 34.3/40.9 & 0.56 -- 0.85\\
2020 Feb 27 & 0841270101 &   58906.31 & 58906.81 & 37.3/45.4 & 0.03 -- 0.35\\
\hline
\end{tabular}
\end{minipage}}
\end{table}

\subsection{Timing analysis results }
\label{sec-timing}

The source pulsations at 13.2 s are cleary visible in the 2018 and 2020 data. The spin periods measured in these new observations are  consistent with those predicted by the ephemeris reported in \citet{mer16}, which were  derived from a phase-coherent timing  of all the data obtained before 2015. Therefore, we included the new data in the phase-coherent analysis. Briefly, this consists in deriving the time of arrivals of a fiducial phase of the pulse profile measured in the different observations and, after correcting for the effect of orbital motion,  fitting them with a quadratic function  $\phi(t) = \phi_0(t) + \nu_0(t-T_0) + 0.5\dot{\nu}(t-T_0)^2$. 
The phases were determined by fitting a sinusoidal function to the 0.15-0.5 keV pulse profiles measured in time intervals of 2000 s.
This  procedure was done iteratively, starting from the most closely spaced observations, and progressively including the other ones, as the improved timing parameters allow to mantain phase coherence. 
The orbital parameters were kept fixed at the values of Table~\ref{tim_par}.

\begin{figure}
\center
\includegraphics[width=9.1cm]{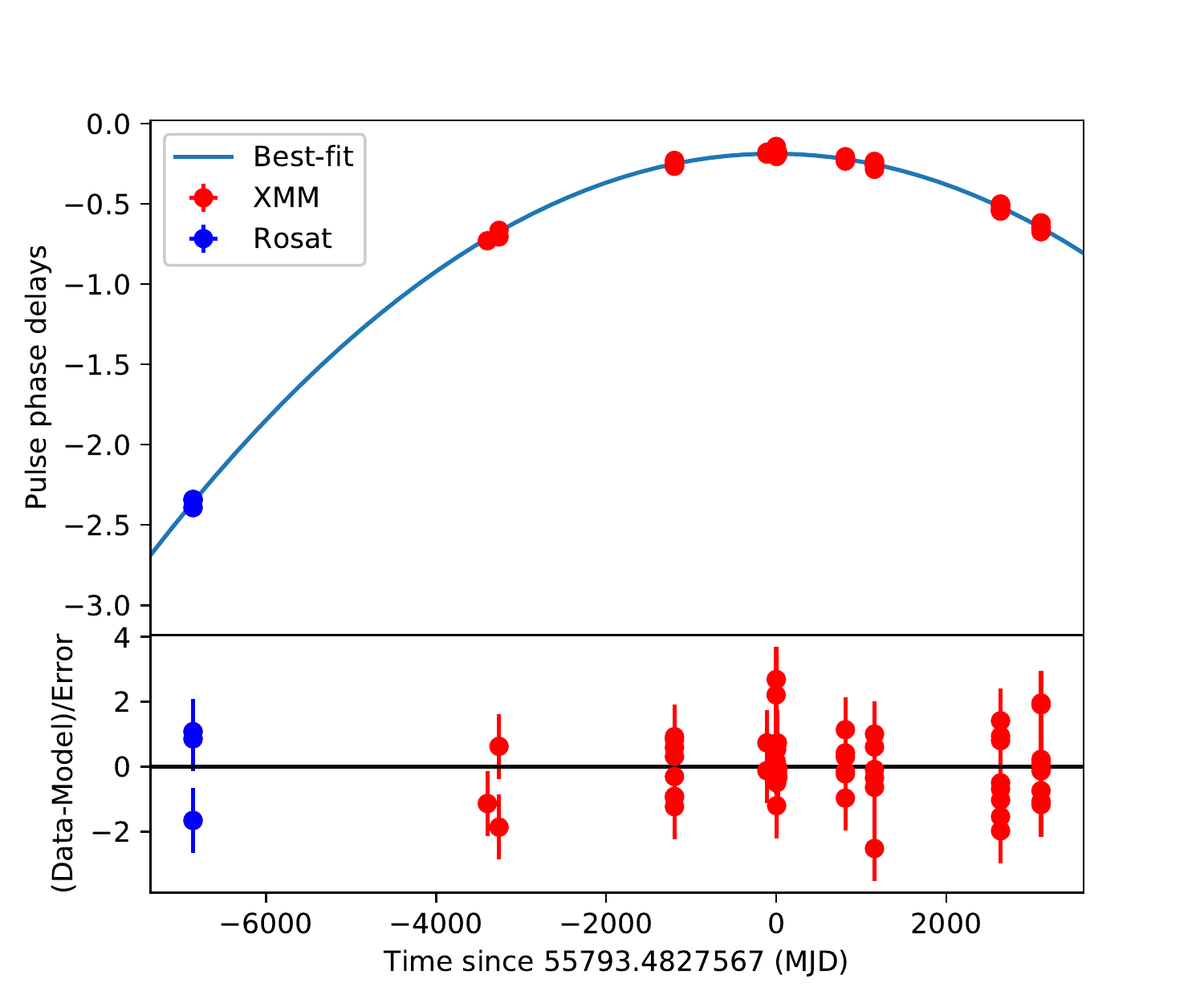}
  \caption{Timing solution (in units of phase, solid blue line) for $\sim28$ years of observations of \hr .   The residuals of the best-fit model are presented in the bottom panel.}  
  \label{tim_sol}
\end{figure}

Our final  solution  provided a good fit  ($\chi_{\nu}^2$ = 1.21 for 44 degrees of freedom (dof)) and a highly significant quadratic term with the parameters given in Table~\ref{tim_par}.
The best fit and its residuals are shown in Fig.~\ref{tim_sol}.  This timing solution is consistent with that of \citet{mer16}, but, thanks to the longer baseline, it has smaller uncertainties on the best fit parameters.

\begin{figure}
\center
\includegraphics[width=8.5cm]{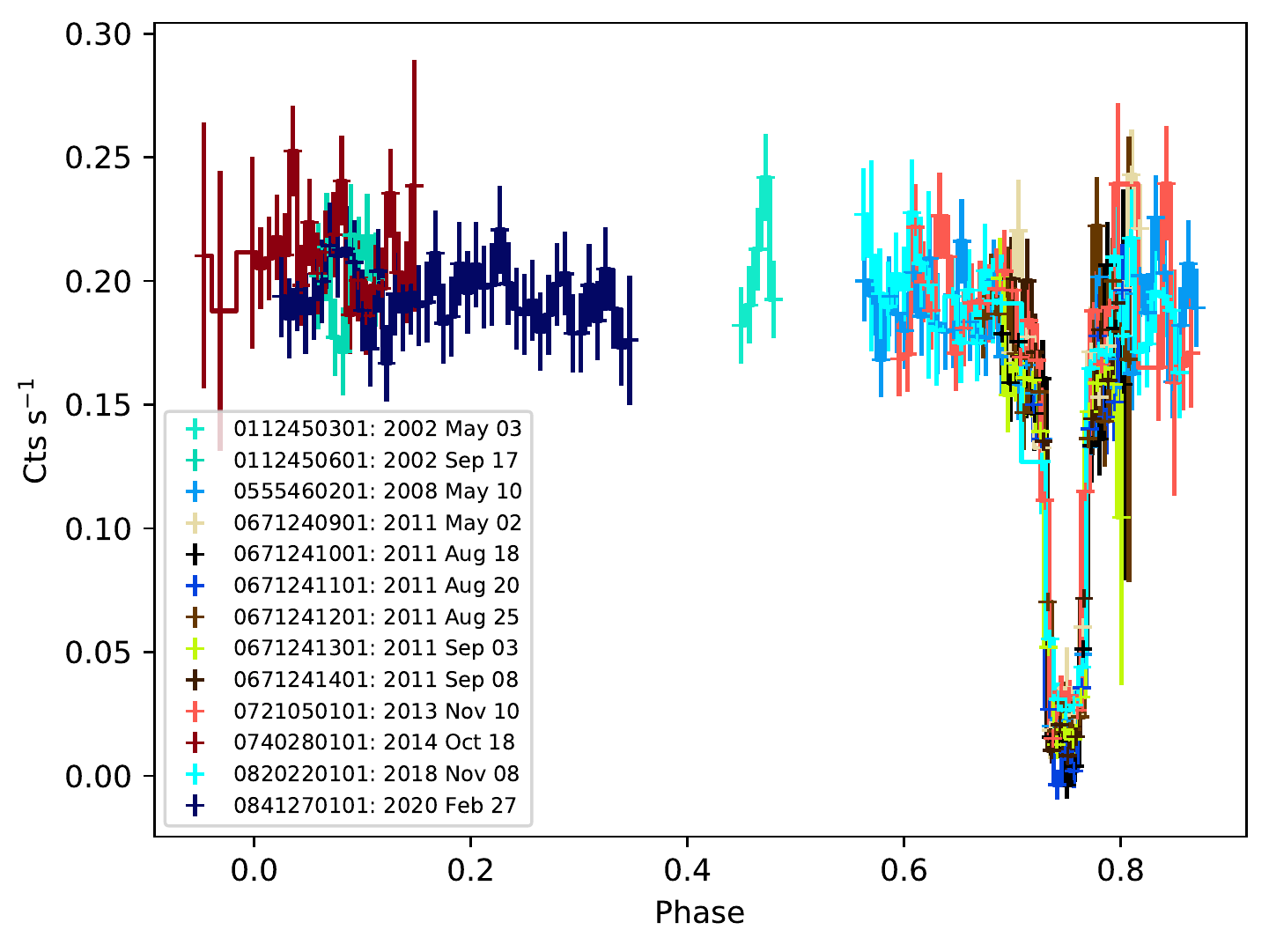}
\includegraphics[width=8.5cm]{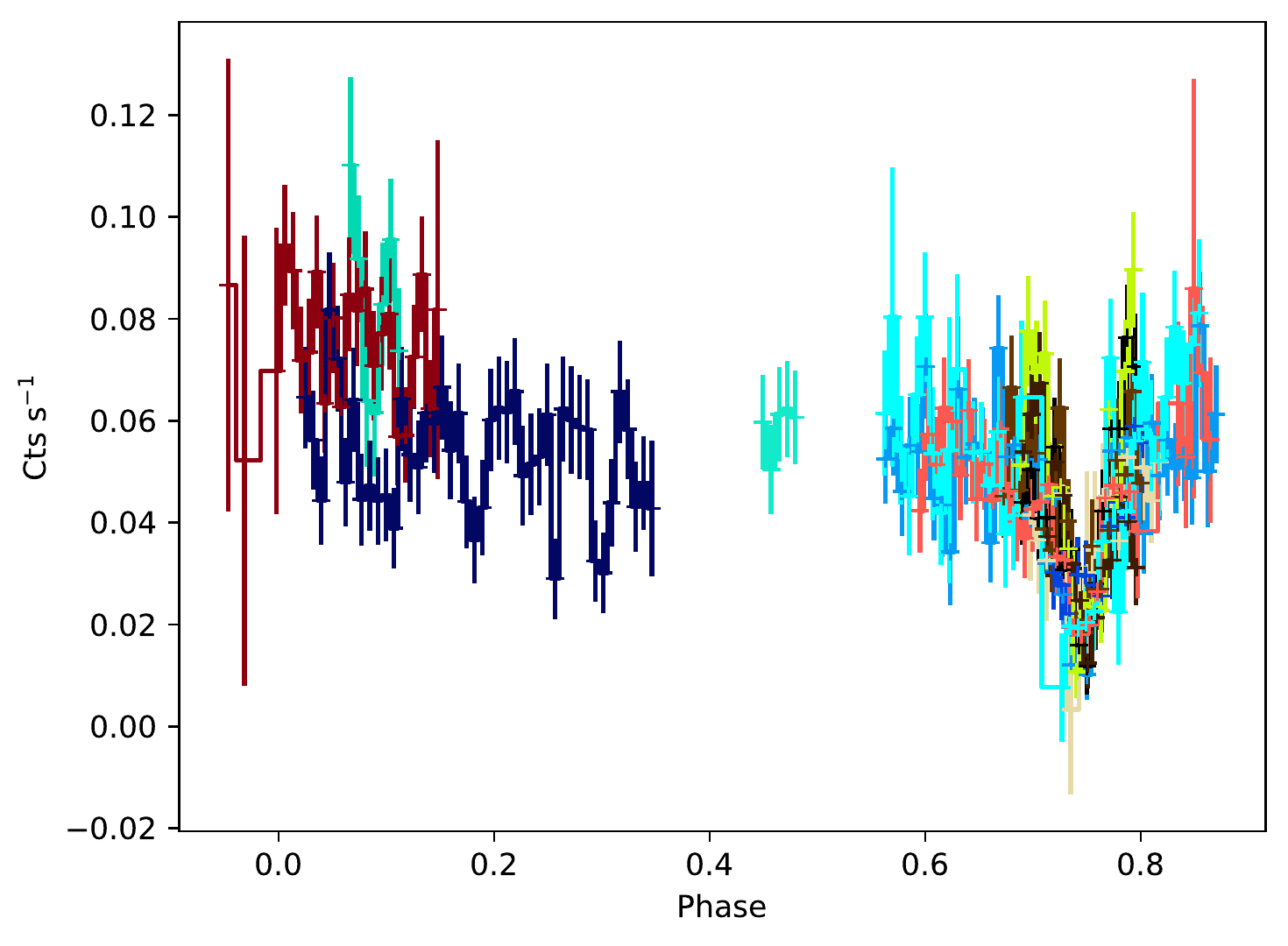}
  \caption{Light curves folded at the orbital period in the soft (0.2--0.5 keV, top) and hard (0.5--10 keV, bottom) energy ranges.
 The eclipse occurs at phase 0.75. The time bin is 1000 s.}
  \label{fig-lc}
\end{figure}

\begin{table}
\caption{Phase-coherent timing solution valid in the 48937.75--58761.81 MJD range.
\label{tim_par}}
\begin{tabular}{lcl}
\hline
 Parameter  &  Value &   Units   \\
\hline
Right Ascension & 6$^h$ 48$^m$ 4.7$^s$  &  J2000  \\
Declination    &   $-44^{\circ}$ 18$'$ 58.4$''$   &  J2000 \\
Orbital period  & 1.547666             &  d      \\ 
A$_X$ sin i    &   9.78646                &  light-s      \\
T$^*$        & 43962.017$^a$            &  MJD \\
$\nu_0$       &     0.075848091873(12) & Hz \\
$\nudot$     &    1.250(7) $\times$ 10$^{-17}$               &  Hz s$^{-1}$    \\
P$_0$          &   13.184247295(2)  & s \\
$\pdot$     &   --2.17(1)  $\times$ 10$^{-15}$          &  s s$^{-1}$ \\
T$_0$    &  55793.4827567 & MJD \\
\hline
\end{tabular}
\begin{small}
\flushleft $^{a}$ Due to a typo, this parameter was incorrectly reported in Table~2 of \citet{mer16}. With this definition of T$^*$ the eclipse occurs at orbital phase 0.75.

\end{small}
\end{table}

To search for long term variability, disentangling the effects related to the orbital phase, we produced  background subtracted light curves folded at the orbital period (Fig.~\ref{fig-lc}). They refer to  the soft (0.2-0.5 keV) and hard (0.5-10 keV) energy bands. It can be seen  that there is some evidence that the hard X-ray  flux in 2020 was slightly lower than in previous observations that covered similar orbital phases.  This is indeed confirmed by the spectral analysis described below.
The 2020 hard X-ray light curve shows also evidence for some variability on timescales of few hundreds seconds. In fact a fit with a constant of the data binned at 400 s yields   $\chi_{\nu}^2$ = 1.66 for 108 dof.

\subsection{X-ray emission during  the eclipse}

The 2018 observation covered the orbital phases around the eclipse of the X-ray pulsar and confirms the presence of significant X-ray emission also when the compact object is occulted by its subdwarf companion, as it was seen in previous eight observations  \citep{mer09,mer13}. To investigate possible long term variability of the emission during the eclipse, we extracted  EPIC-pn images (in  the 0.3--4.5 keV range) in a time interval of  4300 s (the eclipse duration measured in  \citealt{mer13})  centered at phase 0.75   
and carried out source detections using the SAS task {\it edetect\_chain}. \hd\ was detected in all the nine images, with count rates   consistent with a constant value of $0.048 \pm 0.002$ counts s$^{-1}$ ($\chi_{\nu}^2$ = 1.44 for 8 dof).

We then produced a pn and a MOS spectrum of the eclipse emission  by stacking   all the  source spectra of the individual observations, extracted from the above orbital phase interval. These two spectra were  fit simultaneously using the sum of three thermal plasma emission models (vapec in XSPEC) with abundances fixed to the most recent values found for \hd\ (see Table 2 of \citealt{krt19}). The resulting temperatures of 0.15, 0.9, and 6 keV, and the corresponding model normalizations,  were consistent with those found by  \citet{mer16r}.
Under the assumption that the X-ray flux seen during the eclipse is due to intrinsic emission from \hd , and thus present at all orbital phases, we  included this best fit eclipse model  as  a fixed component   in all the subsequent spectral fits discussed below.

\subsection{Variability in the hard X-ray component}

The  2020 observation provided the first data set with a long exposure time at orbital phases far from the eclipse,  which were covered only marginally  by previous \xmm\ pointings. 
For a first analysis of these data and to facilitate the comparison with previous results, we adopted a blackbody plus power law model and we concentrated on a comparison with the 2014 observation, which covered similar orbital phases.

Separate fits to the 2014 and 2020 spectra, in the 0.3-10 keV range, gave consistent values for the blackbody component,  but  different best fit parameters for the power law, indicating a lower flux in  2020.  Therefore, we fitted  the 2014 and 2020 pn spectra simultaneously, imposing common values for all the parameters except for the  power law normalization.   
This resulted in a good fit ($\chi^2_{\nu}=1.15$ for 167 dof) 
with 0.3-10 keV fluxes of the power law component of  $(1.7\pm0.2)\times10^{-13}$ erg cm$^{-2}$ s$^{-1}$ in 2014 and $(1.0\pm0.1)\times10^{-13}$ erg cm$^{-2}$ s$^{-1}$ in 2020.
These flux values confirm the evidence for the  variability seen in the hard X-ray light curves  of Fig~\ref{fig-lc}.
All the other  spectral  parameters were  consistent with those obtained from the sum of all the   data taken before 2015  \citep{mer16}, although they had larger uncertainties due to their lower counting statistics.

\subsection{The soft thermal component}

Since with the new observations we found no  evidence for long term variability in the soft thermal component,  we performed a spectral analysis using the spectra obtained  by stacking 
all the available observations. In the following, we present the results obtained with the EPIC-pn spectra\footnote{We checked that similar results were obtained with the MOS.}, corresponding to a total exposure time of 170 ks.  
In addition to fits with the usual blackbody plus power law model, 
we also explored a more physically motivated scenario, replacing the blackbody component with a white dwarf atmosphere model that we computed specifically for this source. 

The effects of an atmosphere on the emerging thermal radiation depend on many parameters, including elemental  composition and  surface gravity,  but it is impossible to constrain all of them with the limited spectral resolution and narrow bandwidth of the current data.  Therefore, we explored only a single model that we computed assuming  a surface gravity of log~$g$ = 9 (appropriate for this massive white dwarf) and a composition  based on the abundances measured for  \hd\  \citep{krt19}.  This choice for the abundances is based on the hypothesis that the white dwarf surface, or at least the regions where matter accretes and that are responsible for the X-ray emission, are covered by matter coming from the companion star.

\begin{figure}
\center
   \includegraphics[width=9cm, angle=0]{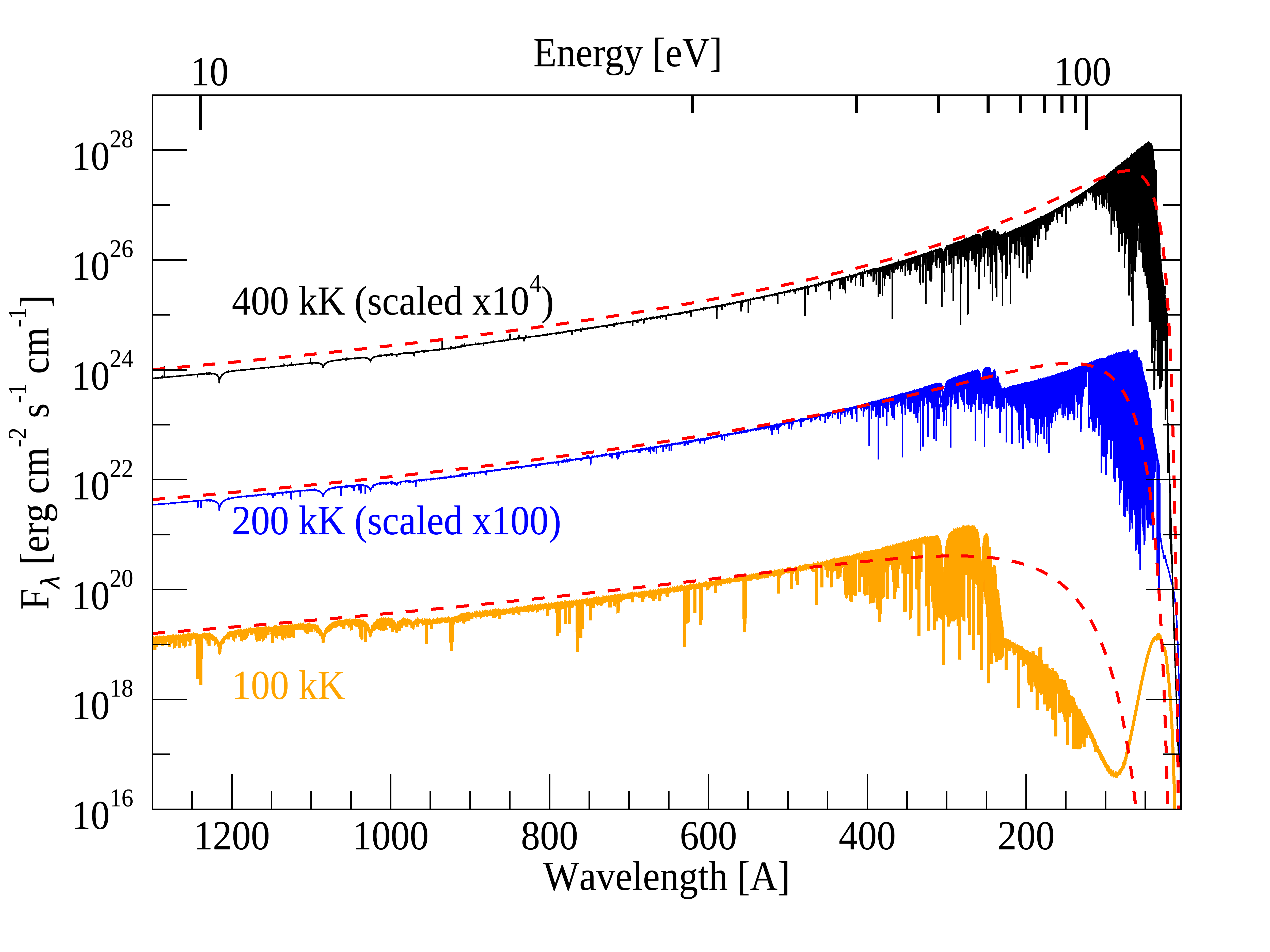}
\caption{Examples of our white dwarf atmosphere model for effective temperatures of 10$^5$, 2$\times10^5$ and  4$\times10^5$ K.   Note that the two latter have been rescaled by two and four decades for clarity. The red dashed lines indicate for comparison blackbody models with the corresponding temperatures.}  
 \label{atm-lambda}
\end{figure}

\begin{figure}
\center
   \includegraphics[width=9cm, angle=0]{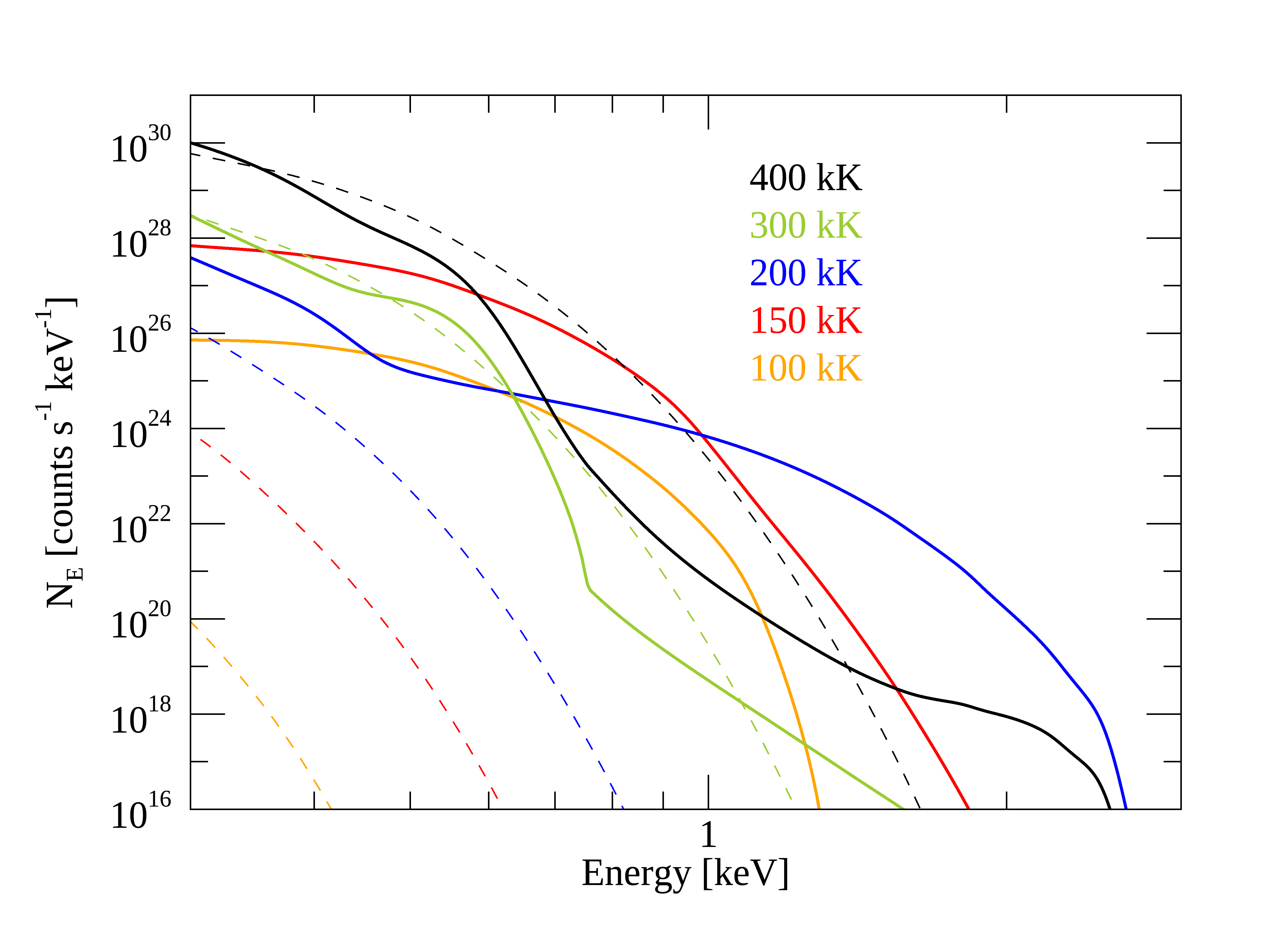}
\caption{Examples of our white dwarf atmosphere model in the 0.3-3 keV energy range folded through the EPIC pn response.    The dashed lines represent blackbody models with the corresponding temperatures.}  
 \label{atm-ene}
\end{figure}

To model stellar spectra in the relevant $T_\mathrm{eff}$ and $\log g$ ranges, 
non-local thermodynamic equilibrium (NLTE) model atmospheres are mandatory. 
Thus, we employed the T\"ubingen NLTE Model-Atmosphere Package
\citep[TMAP\footnote{\url{http://astro.uni-tuebingen.de/~TMAP}},][]{werneretal2003,tmap2012}
to calculate plane-parallel and chemically homogeneous atmospheres in hydrostatic and radiative
equilibrium. We considered opacities of H, He, C, N, O, Fe, and Ni. The atomic data were taken
from the T\"ubingen Model-Atom Database (TMAD) and, for Fe and Ni, calculated with our Iron Opacity and Interface program (IrOnIc), that uses a statistical approach with so-called super levels and super lines \citep{rauchdeetjen2003}
based on Kurucz's line lists\footnote{\url{http://kurucz.harvard.edu/atoms.html}}
\citep{kurucz2009,kurucz2011} and Opacity Project
data\footnote{\url{http://cdsweb.u-strasbg.fr/topbase/topbase.html}} \citep{seatonetal1994}.
We adopted the  abundances given by the following normalized mass fractions:
H  (8.045$\times10^{-2}$),
He (9.134$\times10^{-1}$),
C  (2.337$\times10^{-4}$),
N  (3.431$\times10^{-3}$),
O  (1.239$\times10^{-4}$),
Fe (2.168$\times10^{-3}$), and
Ni (1.817$\times10^{-4}$).
Some examples of the model for different values of the effective temperature are shown in Fig.~\ref{atm-lambda}.  
Due to the presence of many absorption lines and edges, they deviate significantly from blackbody spectra, especially at the shortest wavelengths.  
This is also illustrated in  Fig.~\ref{atm-ene}, which presents an enlarged view of the energy range covered by our X-ray data. In this figure, the theoretical model has been convolved with the instrumental response of the EPIC pn detector,  which results in a smearing of the sharp spectral features.

By fitting the pn spectrum with the atmosphere model plus a power law, we could not obtain a unique best fit solution. In fact, exploring the whole range of temperatures covered by our model\footnote{We imposed values of N$_H>10^{20}$ cm$^{-2}$ to avoid convergence of the fits to unreasonable values.}, we found similarly good results with effective temperatures  of  about 2.2 or 2.9 $\times$10$^5$ K  ($\chi_{\nu}^2$=1.26 and 1.28, respectively).  
These fits imply emitting regions more than one hundred times larger than that derived with the blackbody model  (see Table~\ref{tab-fit}). Another reasonably good fit, with only a slightly worse $\chi_{\nu}^2$=1.39,  was  found for T$_{\rm eff}$=3.7$\times$10$^5$ K.  

The reason for these multiple solutions can be qualitatively understood by examining Fig.~\ref{atm-ene}:    in the $\sim$0.3-0.6 keV range, where the thermal emission dominates over the power law component, the atmosphere model has an average spectral slope that varies with temperature in a non monotonic way. The slope observed in the X-ray data,  and  well described by a blackbody of $\sim3.7\times10^5$ K, can be well approximated by atmosphere models with different values of T$_{\rm eff}$, also considering the effect of changes in N$_H$ and in the power law parameters  (see best fit values in Table~\ref{tab-fit}).

\begin{figure}
\center
    \includegraphics[width=8cm, angle=0]{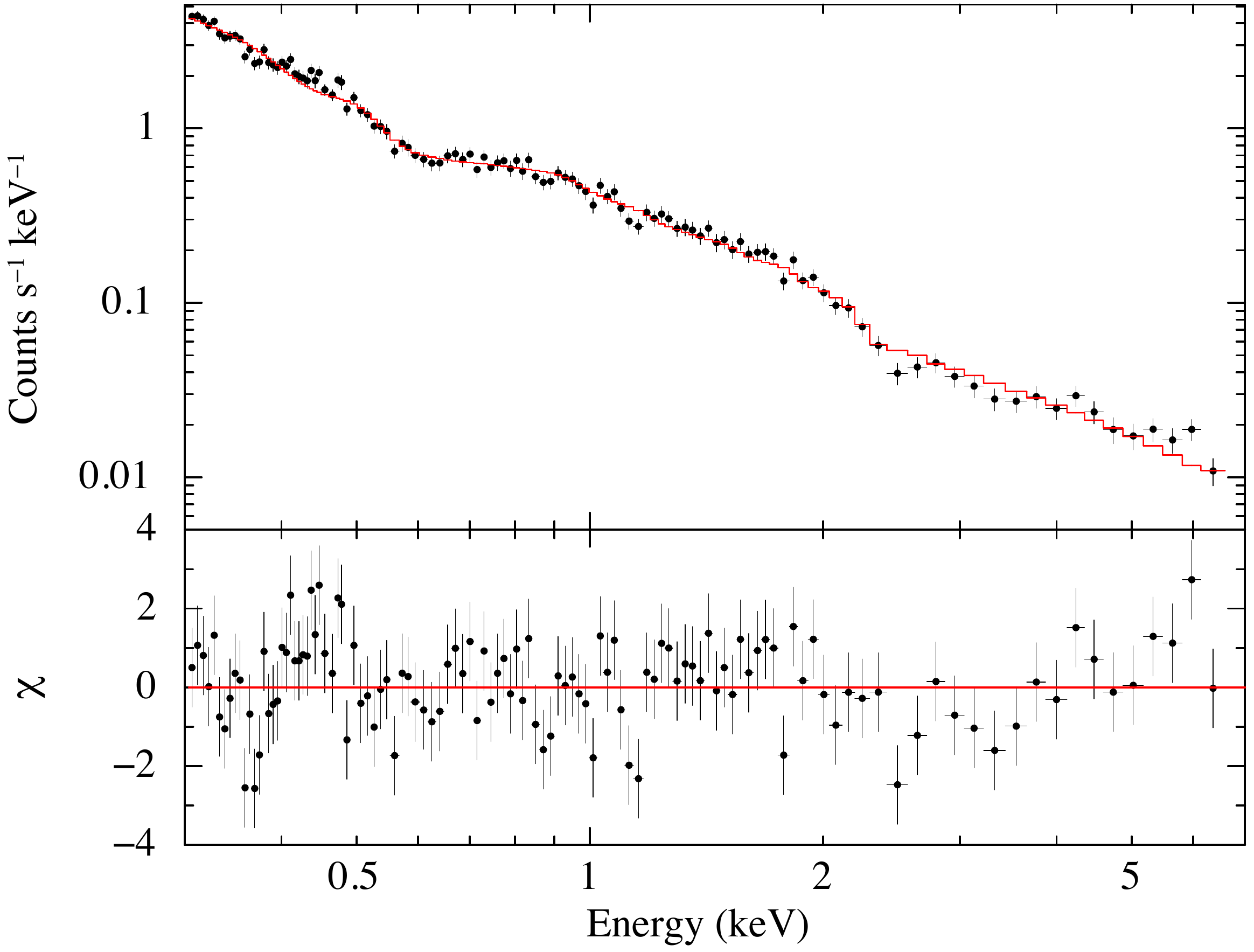}
\caption{Best fit (top) and residuals (bottom) of the total pn spectrum with  a  power law plus atmosphere model with  T$_{\rm eff}$=2.25$\times$10$^5$ K.}  
 \label{atm-225}
\end{figure}

The best fit to the pn spectrum with an atmosphere of T$_{\rm eff}$=2.25$\times$10$^5$ K is shown in Fig.~\ref{atm-225}. The residuals in the fits, which lead to a $\chi_{\nu}^2$ value slightly higher than that obtained with a  blackbody,  could be reduced by changing the atmosphere metal abundances and/or surface gravity  (as well as by allowing some small variations in the eclipse model used to describe the contribution from \hd ). However, besides increasing the number of free parameters, this would not lead to a unique solution and we believe that such more refined analysis should wait for the availability of future data with better spectral resolution and statistics.

\begin{table*} 
\caption{Spectral results   (errors at 90\% c.l.).
\label{tab-fit}
}
\begin{center}
\begin{tabular}{lcccccc}
\hline\\
Model  & N$_H$                     & T$_{\rm eff}$    &                        R$^a$                  & photon index & F$_{\rm PL}^b$        & $\chi_{\nu}^2$/dof \\
         & 10$^{20}$ cm$^{-2}$ & 10$^5$ K           &       km                  &                   &   10$^{-13}$ erg cm$^{-2}$ s$^{-1}$ &    \\
\hline\\
BB+PL  & 1 (fixed)             & 3.65$\pm$0.18            &    41$_{-10}^{+13}$       &1.80$\pm$0.06  &  1.18$\pm$0.05 &1.16/116 \\
\hline\\
Atm+PL& 14.3$\pm$1.5  & 2.25$_{-0.04}^{+0.02}$  & 1604$_{-212}^{+207}$ & 1.75$\pm$0.09 &  1.23$\pm$0.06 &  1.26/116 \\
Atm+PL& 8.5$\pm$0.8    & 2.94$\pm$0.01              &   397$\pm$32               & 1.66$\pm$0.09 &  1.20$\pm$0.06 &  1.28/116 \\
Atm+PL& 6.3$\pm$1.5   &  3.73$_{-0.19}^{+0.04}$  &  66$_{-9}^{+26}$          & 1.92$\pm$0.06 &  1.22$\pm$0.05 &  1.39/116 \\
\hline
\end{tabular}
\flushleft  $^a$ Emission radius for d=521 pc.
\flushleft  $^b$ Unabsorbed flux of power law component in the 0.3--10 keV energy range.
\end{center}
\end{table*}

\section{Discussion and Conclusions}

Evidence for small ($\sim$30\%) variability in the hard X-ray component of \hd\  was first reported by \citet{mer11},   based on the observations of September 2002 and May 2008. Since those data were obtained at very different orbital phases, it was not clear if such variation was related to the source position in the orbit or to real long term changes in the properties of the system. 
The latter possibility seems favoured by our new results, which show a $\sim$70\%  flux variation between two observations spaced by  six years, but  taken at similar orbital phases.  
This suggests that the stellar wind from the hot subdwarf \hd , despite being much weaker than those of normal early type stars, might be subject to  time variable inhomogeneities leading to changes in the mass accretion rate, similar to those seen in high mass X-ray binaries.  The resulting luminosity variations are more prominently appearing in the harder spectral component, likely originating in shocks occurring within the accretion stream.

The new \xmm\ data reported here show that  the pulsar in \hr\ has continued also after 2014 its regular spin-up at the same rate of   $\pdot = -2.17\times10^{-15}$ s s$^{-1}$. 
As discussed in \citet{mer16}, the mass accretion rate in this system is insufficient to cause this rapid spin-up if the compact object is a  white dwarf. On the other hand,  such a remarkably  steady spin-up  rate is difficult to explain for a neutron star subject to wind accretion:  changes in the torque, resulting from variations in the stellar wind gravitationally captured by the compact object, should affect its spin period. However, no changes in   $\pdot$ were seen,  despite the observed long term variability.
These   results support the intepretation of the spin-up in terms of the secular contraction of a   white dwarf of few Myr age, as proposed by \citet{pop18}.

Most of the X-ray luminosity of \rx\ is emitted  in a soft thermal component, probably originating on the white dwarf surface. The large size of the emitting area,  inferred from   blackbody fits, has been  one of the arguments used to disfavour a neutron star interpretation. 
Here we presented the first attempt to describe this soft component with  a more physical model, accounting for the presence of an atmosphere 
and assuming  that the white dwarf is covered by helium-rich material accreted from its companion star.
Fitting with this model plus a power law, we obtained reasonably good results, although the temperature could not be uniquely  constrained.  
Remarkably, these   fits   result in  significantly larger emission radii than those derived with the blackbody model, implying that the thermal component originates  from a large fraction, or even from the totality, of the star surface. 

Another relevant implication of these fits is that the contribution of the thermal emission in the  UV/optical bands is much larger than that expected by the  extrapolation of the best fit blackbody model.  For example, the best atmosphere fit with   T$_{\rm eff}$=2.25$\times$10$^5$ K gives an optical flux about three orders of magnitude larger than that of the     blackbody.  We cannot exclude that  models with slightly different compositions could give an even larger flux at long wavelengths, when fitted to the X-ray data.   This emission might produce detectable pulsations in the optical band,  if it is modulated as strongly as in the soft X-ray band.  Previous searches for optical pulsations at the spin period of 13.2 s provided an upper limit  of    6$\times$10$^{-4}$ photons cm$^{-2}$ s$^{-1}$  \AA$^{-1}$ at  3600 \AA\  \citep{mer11}.  The extrapolation of the  best fit atmosphere model is only about one order of magnitude below this upper limit, suggesting that more sensitive searches  could lead to the detection of optical pulsations or to useful constraints in case of null results.

\section{Summary}

Thanks to new \xmm\ observations of \hr\ carried out in 2018 and 2020, we could extend its phase-connected timing solution,  now spanning a time interval of almost thirty years, without finding any evidence of variations in the spin-up rate, despite the long term variability in the flux of the hard  X-ray component reported here for the first time. 

We computed a specific white dwarf atmosphere model for this system, adopting  appropriate composition and surface gravity values, and used it to describe the soft X-ray component, that was traditionally fitted with a blackbody.  Our analysis showed  that,  in order to fully exploit the potential diagnostics  of physical models of this type,  data with higher spectral resolution are needed, as can be provided by future experiments like  XRISM \citep{XRI20} and Athena/XIFU \citep{bar18short}. 
On the other hand, this first attempt indicates that  atmosphere models lead to  lower temperatures   and much larger emission regions compared to blackbody fits. Interestingly, our best fit results  imply that the pulsed thermal component emitted from the compact object gives a contribution in the optical/UV bands  much larger than predicted by the blackbody. This opens  promising prospects for future searches of  optical pulsations and, more in general, for the investigation of  variability in the optical band, where the contribution from the bright and hot \hd\ is dominant.

\section*{Acknowledgements}

We acknowledge financial support from INAF, through  grant DP n.43/18 (``Main-streams''), and from the Italian Ministry for University and Research, through grant 2017LJ39LM (``UNIAM''). The TMAD service (\url{http://astro-uni-tuebingen.de/~TMAD}) used for this paper
was constructed as part of the activities of the German Astrophysical Virtual Observatory.
IP and VS were partially funded by the Deutsche Forschungsgemeinschaft (DFG) under grants GE2506/12-1 and GE2506/9-1, respectively.
 This work is based on data from observations with {\it XMM-Newton}, an ESA science mission with instruments and contributions directly funded by ESA Member States and the USA (NASA). 

\section*{Data Availability}
The data underlying this article will be shared on reasonable request to the corresponding author.

\addcontentsline{toc}{section}{Bibliography}
\bibliographystyle{mn2e}

\begin{thebibliography}{}

\bibitem[\protect\citeauthoryear{{Barret}, {Lam Trong} \& {den
  Herder}}{{Barret} et~al.}{2018}]{bar18short}
{Barret} D.,  {Lam Trong} T.,    {den Herder} e.~a.,  2018, in {den Herder}
  J.-W.~A.,  {Nikzad} S.,   {Nakazawa} K.,  eds, Space Telescopes and
  Instrumentation 2018: Ultraviolet to Gamma Ray Vol. 10699 of Society of
  Photo-Optical Instrumentation Engineers (SPIE) Conference Series, {The ATHENA
  X-ray Integral Field Unit (X-IFU)}.
p. 106991G

\bibitem[\protect\citeauthoryear{{Brooks}, {Kupfer} \& {Bildsten}}{{Brooks}
  et~al.}{2017}]{bro17}
{Brooks} J.,  {Kupfer} T.,    {Bildsten} L.,  2017, \apj, 847, 78

\bibitem[\protect\citeauthoryear{Brown, Vallenari, Prusti, de Bruijne,
  Babusiaux \& Biermann}{Brown et~al.}{2020}]{gaia20}
Brown A. G.~A.,  Vallenari A.,  Prusti T.,  de Bruijne J. H.~J.,  Babusiaux C.,
     Biermann M., , 2020, Gaia Early Data Release 3: Summary of the contents
  and survey properties

\bibitem[\protect\citeauthoryear{{Hamann}, {Gruschinske}, {Kudritzki} \&
  {Simon}}{{Hamann} et~al.}{1981}]{ham81}
{Hamann} W.,  {Gruschinske} J.,  {Kudritzki} R.~P.,    {Simon} K.~P.,  1981,
  \aap, 104, 249

\bibitem[\protect\citeauthoryear{{Hamann}}{{Hamann}}{2010}]{ham10}
{Hamann} W.-R.,  2010, \apss, 329, 151

\bibitem[\protect\citeauthoryear{{Heber}}{{Heber}}{2016}]{heb16}
{Heber} U.,  2016, \pasp, 128, 082001

\bibitem[\protect\citeauthoryear{{Israel}, {Stella}, {Angelini}, {White},
  {Kallman}, {Giommi} \& {Treves}}{{Israel} et~al.}{1997}]{isr97}
{Israel} G.~L.,  {Stella} L.,  {Angelini} L.,  {White} N.~E.,  {Kallman} T.~R.,
   {Giommi} P.,    {Treves} A.,  1997, \apjl, 474, L53

\bibitem[\protect\citeauthoryear{{Krti{\v{c}}ka}, {Jan{\'\i}k},
  {Krti{\v{c}}kov{\'a}}, {Mereghetti}, {Pintore}, {N{\'e}meth}, {Kub{\'a}t} \&
  {Vu{\v{c}}kovi{\'c}}}{{Krti{\v{c}}ka} et~al.}{2019}]{krt19}
{Krti{\v{c}}ka} J.,  {Jan{\'\i}k} J.,  {Krti{\v{c}}kov{\'a}} I.,  {Mereghetti}
  S.,  {Pintore} F.,  {N{\'e}meth} P.,  {Kub{\'a}t} J.,    {Vu{\v{c}}kovi{\'c}}
  M.,  2019, \aap, 631, A75

\bibitem[\protect\citeauthoryear{{Kudritzki} \& {Simon}}{{Kudritzki} \&
  {Simon}}{1978}]{kud78}
{Kudritzki} R.~P.,  {Simon} K.~P.,  1978, \aap, 70, 653

\bibitem[\protect\citeauthoryear{{Kurucz}}{{Kurucz}}{2009}]{kurucz2009}
{Kurucz} R.~L.,  2009, in {Hubeny} I.,  {Stone} J.~M.,  {MacGregor} K.,
  {Werner} K.,  eds, American Institute of Physics Conference Series Vol.~1171
  of American Institute of Physics Conference Series, {Including All the
  Lines}.
pp 43--51

\bibitem[\protect\citeauthoryear{{Kurucz}}{{Kurucz}}{2011}]{kurucz2011}
{Kurucz} R.~L.,  2011, Canadian Journal of Physics, 89, 417

\bibitem[\protect\citeauthoryear{{Mereghetti} \& {La Palombara}}{{Mereghetti}
  \& {La Palombara}}{2016}]{mer16r}
{Mereghetti} S.,  {La Palombara} N.,  2016, Advances in Space Research, 58, 809

\bibitem[\protect\citeauthoryear{{Mereghetti}, {La Palombara}, {Tiengo},
  {Pizzolato}, {Esposito}, {Woudt}, {Israel} \& {Stella}}{{Mereghetti}
  et~al.}{2011}]{mer11}
{Mereghetti} S.,  {La Palombara} N.,  {Tiengo} A.,  {Pizzolato} F.,  {Esposito}
  P.,  {Woudt} P.~A.,  {Israel} G.~L.,    {Stella} L.,  2011, \apj, 737, 51

\bibitem[\protect\citeauthoryear{{Mereghetti}, {La Palombara}, {Tiengo},
  {Sartore}, {Esposito}, {Israel} \& {Stella}}{{Mereghetti}
  et~al.}{2013}]{mer13}
{Mereghetti} S.,  {La Palombara} N.,  {Tiengo} A.,  {Sartore} N.,  {Esposito}
  P.,  {Israel} G.~L.,    {Stella} L.,  2013, \aap, 553, A46

\bibitem[\protect\citeauthoryear{{Mereghetti}, {Pintore}, {Esposito}, {La
  Palombara}, {Tiengo}, {Israel} \& {Stella}}{{Mereghetti}
  et~al.}{2016}]{mer16}
{Mereghetti} S.,  {Pintore} F.,  {Esposito} P.,  {La Palombara} N.,  {Tiengo}
  A.,  {Israel} G.~L.,    {Stella} L.,  2016, \mnras, 458, 3523

\bibitem[\protect\citeauthoryear{{Mereghetti}, {Tiengo}, {Esposito}, {La
  Palombara}, {Israel} \& {Stella}}{{Mereghetti} et~al.}{2009}]{mer09}
{Mereghetti} S.,  {Tiengo} A.,  {Esposito} P.,  {La Palombara} N.,  {Israel}
  G.~L.,    {Stella} L.,  2009, Science, 325, 1222

\bibitem[\protect\citeauthoryear{{Popov}, {Mereghetti}, {Blinnikov}, {Kuranov}
  \& {Yungelson}}{{Popov} et~al.}{2018}]{pop18}
{Popov} S.~B.,  {Mereghetti} S.,  {Blinnikov} S.~I.,  {Kuranov} A.~G.,
  {Yungelson} L.~R.,  2018, \mnras, 474, 2750

\bibitem[\protect\citeauthoryear{{Rauch} \& {Deetjen}}{{Rauch} \&
  {Deetjen}}{2003}]{rauchdeetjen2003}
{Rauch} T.,  {Deetjen} J.~L.,  2003, in {Hubeny} I.,  {Mihalas} D.,   {Werner}
  K.,  eds, Stellar Atmosphere Modeling Vol.~288 of Astronomical Society of the
  Pacific Conference Series, {Handling of Atomic Data}.
p.~103

\bibitem[\protect\citeauthoryear{{Seaton}, {Yan}, {Mihalas} \&
  {Pradhan}}{{Seaton} et~al.}{1994}]{seatonetal1994}
{Seaton} M.~J.,  {Yan} Y.,  {Mihalas} D.,    {Pradhan} A.~K.,  1994, \mnras,
  266, 805

\bibitem[\protect\citeauthoryear{{Str{\"u}der}, {Briel}, {Dennerl} \& {et
  al.}}{{Str{\"u}der} et~al.}{2001}]{str01}
{Str{\"u}der} L.,  {Briel} U.,  {Dennerl} K.,    {et al.} 2001, \aap, 365, L18

\bibitem[\protect\citeauthoryear{{Thackeray}}{{Thackeray}}{1970}]{tha70}
{Thackeray} A.~D.,  1970, \mnras, 150, 215

\bibitem[\protect\citeauthoryear{{Turner}, {Abbey}, {Arnaud} \& {et
  al.}}{{Turner} et~al.}{2001}]{tur01}
{Turner} M.~J.~L.,  {Abbey} A.,  {Arnaud} M.,    {et al.} 2001, \aap, 365, L27

\bibitem[\protect\citeauthoryear{{Wang} \& {Han}}{{Wang} \&
  {Han}}{2010}]{wan10}
{Wang} B.,  {Han} Z.-W.,  2010, Research in Astronomy and Astrophysics, 10, 681

\bibitem[\protect\citeauthoryear{{Werner}, {Deetjen}, {Dreizler}, {Nagel},
  {Rauch} \& {Schuh}}{{Werner} et~al.}{2003}]{werneretal2003}
{Werner} K.,  {Deetjen} J.~L.,  {Dreizler} S.,  {Nagel} T.,  {Rauch} T.,
  {Schuh} S.~L.,  2003, in {Hubeny} I.,  {Mihalas} D.,   {Werner} K.,  eds,
  Stellar Atmosphere Modeling Vol.~288 of Astronomical Society of the Pacific
  Conference Series, {Model Photospheres with Accelerated Lambda Iteration}.
p.~31

\bibitem[\protect\citeauthoryear{{Werner}, {Dreizler} \& {Rauch}}{{Werner}
  et~al.}{2012}]{tmap2012}
{Werner} K.,  {Dreizler} S.,    {Rauch} T., , 2012, {TMAP: T{\"u}bingen NLTE
  Model-Atmosphere Package}, Astrophysics Source Code Library [record
  ascl:1212.015]

\bibitem[\protect\citeauthoryear{{Wu} \& {Wang}}{{Wu} \& {Wang}}{2019}]{wu19}
{Wu} C.,  {Wang} B.,  2019, \mnras, 486, 2977

\bibitem[\protect\citeauthoryear{{Wu}, {Chen}, {Li} \& {Han}}{{Wu}
  et~al.}{2018}]{wu18}
{Wu} Y.,  {Chen} X.,  {Li} Z.,    {Han} Z.,  2018, \aap, 618, A14

\bibitem[\protect\citeauthoryear{{XRISM Science Team}}{{XRISM Science
  Team}}{2020}]{XRI20}
{XRISM Science Team} 2020, arXiv e-prints, p. arXiv:2003.04962

\bibitem[\protect\citeauthoryear{{Yungelson} \& {Tutukov}}{{Yungelson} \&
  {Tutukov}}{2005}]{yun05}
{Yungelson} L.~R.,  {Tutukov} A.~V.,  2005, Astronomy Reports, 49, 871

\end{thebibliography}

\bsp
\label{lastpage}
\end{document}